\documentclass[printer]{aa}
\usepackage[]{graphicx}
\usepackage{natbib}
\def\ltsima{$\; \buildrel < \over \sim \;$}
\def\simlt{\lower.5ex\hbox{\ltsima}}
\def\gtsima{$\; \buildrel > \over \sim \;$}
\def\simgt{\lower.5ex\hbox{\gtsima}}

\def\ergs{{erg s$^{-1}$}}
\def\cm2{{cm$^{-2}$}}

\def\lum{{$L_{2-10\,keV}$}}
\def\p1{{Paper I}}

\def\xmm{{\em XMM--Newton}}
\def\chandra{{\em Chandra}}

\def\ginga{{\em Ginga}}
\def\asca{{\em ASCA}}

\def\nh{{N$_{\rm H}$}}
\def\chandra{{\em Chandra}}

\def\xmm{{\em XMM--Newton}}

\def\nh{{N$_{\rm H}$}}  
\def\epic{{\em EPIC}}

\def\pn{{\em pn}}
\def\mos{{\em MOS}}

\def\f14{{10$^{-14}$}}
\def\f13{{10$^{-13}$}}
\begin{document}

   \title{The XMM--Newton view of PG quasars}

   \subtitle{II. Properties of the Fe K$_\alpha$ line}
\author{E. Jim\'enez-Bail\'on, E. Piconcelli, M. Guainazzi,
N. Schartel, P.M. Rodr\'{\i}guez-Pascual, M. Santos-Lle\'o}
 \institute{XMM--Newton Science Operation Center/RSSD--ESA, Apartado
 50727, E--28080 Madrid, Spain} \authorrunning{XMM~PG~Team}
 \titlerunning{XMM--Newton observation of PG quasars}
 \offprints{ejimenez@xmm.vilspa.esa.es}
 \date{Accepted 21 December 2004}

\abstract{The  properties of the fluorescence Fe  K$_\alpha$ emission
lines of a sample of 38 quasars (QSOs) observed with \xmm~are studied.
These objects are included in  the optically selected sample from  the
Palomar--Green (PG) Bright  Quasar  Survey with  an  X--ray luminosity
$1.3\times     10^{43}<$L$_{2-10\,keV}<5.1\times10^{45}$~\ergs     and
z$\leq$1.72.  For each  object   in the  sample,  we investigated  the
presence of both narrow and broad iron lines in detail.  A total of 20
out of  the 38 QSOs show evidence   of an Fe  K$_\alpha$ emission line
with a narrow profile.  The majority  of the lines are consistent with
an origin in low ionization material, which is likely to be located
in the outer parts  of the accretion  disk, the molecular torus, and/or
the  Broad Line  Region.  The   average  properties of the narrow   Fe
K$_\alpha$ emission  line observed in the  sample are similar to those
of Seyfert type galaxies as inferred from recent \xmm~and
\chandra~studies.  A broad line has  been significantly  detected in
only three objects.   Furthermore,  we   studied  the  relationship  between  the
equivalent width  (EW) of  the  iron line   and the hard   band X--ray
luminosity  for radio  quiet quasars.  The analysis indicates
that  no clear correlation  between the strength  of the  line and the
hard  X--ray luminosity  is present,   and our    results do  not  show  
compelling evidence  for    an   anticorrelation between   these   two
quantities, i.e. the so-called {\it X--ray Baldwin effect}. 

   }

   \maketitle
%

\section{Introduction}
The iron K$_\alpha$ emission line
was one of the first atomic features discovered in  X--ray spectra
of active galaxies \citep{mushotzky}. Its origin was
promptly interpreted as photoelectric absorption followed by
fluorescence emission (see \citet{reynolds} for a 
review).  The combination of fluorescent yield ($\propto Z^4$) and the
(relatively) high abundance of iron justifies the strength of such a
feature with respect to lines emitted by other atomic species. The
K$_\alpha$ line  actually consists of two different lines that correspond
to the 2$p$$\rightarrow$1$s$ transition and that, in the case of
`neutral' iron
(i.e. FeI--XVII), have an energy of 6.404 and 6.391 keV, respectively.

However, the resolving power of earlier X--ray spectrometers does not
allow   for separating both   features  (but see  also  \citet{kaspi} for
results  with   \chandra~HETGS), so  they   are usually  observed (and
considered)  as a single  emission feature.  The maximum energy  for
this line is 6.97 keV corresponding to emission from H--like iron.

The X--ray spectral survey of Seyfert galaxies carried out with {\it
Ginga} \citep{nandrapounds} found this feature in the
vast majority of these sources.  In the same years, several
theoretical works suggested an origin for this feature in
circumnuclear optically--thick material (i.e. the disk and/or the molecular torus),
together with  X--ray Compton 
reflection \citep{georgefabian,matt,krolik,ghisellini94}

The   study of   the iron  K$_\alpha$    emission line  became one  of
priorities of X--ray astronomy after the discovery  of a line with
a broadened and  skewed profile in  the \asca~spectrum of MCG--6-30-15
and\citep{tanaka}.     Such  a broad  profile  was    interpreted as  the
combination  of  transverse  Doppler shift,  relativistic beaming,  and
gravitational redshift. It was the  first direct evidence of  emission
from an accretion disk extending down to  very few Schwarzschild radii
from a supermassive blackhole. Iron lines are therefore a unique tool
for studying strong  gravity effects and  mapping the innermost regions of the
accretion flow.

The advent      of  \xmm~and  \chandra~has   allowed   more   detailed
investigations of this spectral  feature and provided important insights
into its properties. In particular, \citet{yaqoob} and \citet{pounds}
pointed out the ubiquity of an unresolved Fe K$_\alpha$ line in the spectra
of bright Seyfert galaxies.  Such a  line should therefore emerge from
the outermost     regions   of the  accretion  disk,     but non--disk
contributions \citep[i.e. from the Broad  Line Region][]{yaqoob01} are
also plausible.   Contrary  to  previous  low--resolution  {\it  ASCA}
observations \citep{nandra} iron  emission lines with a broad  profile
appear to be quite rare in \chandra~and \xmm~observations.\\

It is  worth noting that  the Fe K$_\alpha$ line has not
commonly been observed in the X--ray  spectra of high luminosity AGNs
(i.e.  QSOs) and, when detected, it is  usually weaker than in Seyfert
galaxies \citep{reeves}.  To explain this effect, \citet{iwasawa}
suggested, based on {\it Ginga} data, the existence of an X--ray
``Baldwin  effect'' such that, as the   X--ray luminosity increases, the
upper layers  of the disk become progressively  more ionized until any
reprocessed feature    is  suppressed \citep{ross}.   The X--ray
Baldwin effect was also  observed with {\it  ASCA} (Nandra et al.
1997) and more recently with {\it XMM--Newton}  (Page et al.  2004a).
However, due to  selection  effects (i.e.   relative  faintness, lower
density  surface  than in Seyfert   galaxies)   and to the limited  spectral
capabilities of past   X--ray   telescopes,  knowledge   about   the
properties of the iron emission line in QSOs is quite sparse.\\

Our extensive \xmm~program \citep{pgpaper1} is designed to fill this lack of
information by detailed and uniform analysis of 40 \epic~spectra
of optically--bright QSOs selected from the Palomar--Green survey
\citep{pgs}.  With such a large sample we aim to provide
useful insights into the characteristics of the Fe K$_\alpha$ line and
its origin. Furthermore, the homogeneity of our sample strongly prevents
possible biases due to mixing different classes of AGNs (see Sect.~\ref{sect:discussion}).

\section{\xmm~observation}

This  study presents an  analysis of  the iron K  band  of all public
observations of PG  quasars available   in the \xmm~Science   Archive
(XSA) as  of  February 2004. Out of the 42 original objects, two
(0003$+$199 and 1426$+$015) were excluded due to large pile--up and
two others (1001$+$054 and 1404$+$226) due to low statistics in the
hard X--ray band, i.e. $>$2~keV. The spectra were obtained by processing the
\xmm~observations using the Science Analysis System (SAS)
v.5.4.1. Sample properties and data reduction are described in detail in \citet{pgpaper1}. {\it
MOS1} and {\it MOS2} 0.6-10~keV spectra were combined and
analyzed simultaneously with 0.3-12~keV {\it pn} spectrum.

\section{Analysis of spectra}

\subsection{Best fit models}

The  spectrum  of  each  QSO was analyzed  using  XSPEC~v.11.2,
\citep{arnaud}.  From \citet{piconcelli}, we
obtained  the best fit  model for the continuum  emission  of each QSO
studied .  

Table~\ref{tab:iron}  shows the  best   fit  model obtained  for  each
source.  A complete description of   these models and their parameters
are presented in \citet{pgpaper1}.  In the following, all the energies
of the QSOs are referred to the rest frame,  and errors are given to the
90 per cent  confidence level  (i.e.   $\Delta\chi^2$  = 2.71)  unless
otherwise stated. Luminosities were calculated considering a flat
cosmology  with  ($\Omega_{\rm M}$,$\Omega_{\rm \Lambda}$) = (0.3,0.7)
and $H_0=70$~km~s$^{-1}$~Mpc$^{-1}$ \citep{bennett}.

\subsection{Spectral models for the Fe line}
\label{sec:iron}

First, we investigate the presence of fluorescent Fe emission lines by
adding  an unresolved  narrow Gaussian line\footnote{The intrinsic
width of the
Gaussian line was forced to be  the
instrumental      resolution   at           the    given       energy,
FWHM$\sim$~7000~km~s$^{-1}$  for  both   \pn~and \mos, $\sigma$=0~keV.}
to    the best  fit  models   reported  in  \citet{pgpaper1}. Whenever the addition of this  component was not required by
the data (i.e. its significance is $<$  95\% according to an $F-$test),
we then  fixed the centroid of the  line to 6.4  keV.  In  such way an
estimate of  upper limit to  the rest--frame equivalent width  (EW) of
the neutral  iron  fluorescence emission was inferred.   The best--fit
values    for    the   emission  line    parameters   are   listed  in
Table~\ref{tab:iron}.  An iron fluorescent emission line with a narrow
profile is significantly  detected, i.e. with an  $F-test$ criterion  at a
significance level of $\geq$95\%, in 20  QSOs ( 17 radio quiet
and 3 radio loud).
\begin{table*}
\caption{Iron line properties. The table includes the name of the QSO,
the best  fit model of the  data, the rest frame energy and  intensity  of the Fe
K$_\alpha$ line modeled  by a unresolved   Gaussian line,  the rest--frame
equivalent  width, the  F--test   probability,   and the  rest--frame
equivalent width of a broad iron line modeled by a {\tt LAOR} line.}
\label{tab:iron}
\begin{footnotesize}\begin{center}
\begin{tabular}{l c c c c c c c }

\hline\hline
\multicolumn{1}{l} {{\bf PG Name }} &
\multicolumn{1}{c} {{\bf Model$^a$} } &
\multicolumn{1}{c} {{\bf L$_{2-10\,keV}$}} &
\multicolumn{4}{c} {{\bf Narrow Line}} &
\multicolumn{1}{c} {{\bf Broad Line}} \\

 & & & $ E_{Fe}$ & EW
  &Intensity&P($F-test$)&EW\\ 
& & (10$^{44}$ erg s$^{-1}$)  &(keV)& (eV) & (ph cm$^{-2}$s$^{-1}$)&(\%)
 &(eV)\\  \hline\hline\\
0007$+$106$^\star$ & A 	     &       1.4     	&6.4f$^{\dag}$&$<$60 &$<$6 $\times$ 10$^{-6}$&$<$75  &220$^{+110}_{-100}$$^{(\dag\dag)}$\\ 
0050$+$124 & E$^\ddag$   &        0.78      & $6.47^{+0.06}_{-0.11}$ & $40\pm30$  & $(4\pm2)\times10^{-6}$ & 97.5 & $100^{+40}_{-50}\,^{(\dag\dag)}$\\
0157$+$001 & A 	     &        0.71      &6.4f &  $<330$   &$<4\times10^{-6}$ 	& $<$75  & $<$1300 \\
0804$+$761 & D 	     &       2.86       &6.38$\pm0.06$ & 100$^{+70}_{-60}$&1.2$^{+0.9}_{-0.7}$$\times$10$^{-5}$&95.7&250$^{+200}_{-180}$\\ 
0844$+$349 & E 	     &        0.55      &6.4f & $<100$                & $<6\times10^{-6}$ 	& $<$75& $<$1200 \\ 
0947$+$396 & C$^\ddag$   &        2.26      &$6.40^{+0.11}_{-0.07}$            &$120\pm60$            & $3.1\pm0.2\times10^{-6}$ 		& 99.8& $<360$\\
0953$+$414 & E$^\ddag$   &       5.41       &6.4f &$<50$&$<4\times$ 10$^{-6}$&$<$75  &$<$77\\
1048$+$342 &  E      &       1.10       &6.34$^{+0.19}_{-0.04}$ &100$\pm60$&1.9$^{+1.1}_{-1.2}$$\times$10$^{-6}$&97.6  &$<$160\\
1100$+$772$^\star$ &  E      &       11.19      &6.40$^{+0.10}_{-0.04}$ &60$\pm40$&3$^{+3}_{-2}$ $\times$ 10$^{-6}$&95  &$<$110\\
1114$+$445 &  G      &       1.45       &6.45$^{+0.02}_{-0.08}$ &100$^{+30}_{-40}$&3.3$^{+1.1}_{-1.2}$$\times$ 10$^{-6}$&99.99  &$<$33\\
1115$+$080 & PL$^\ddag$  &        65.3      & 6.4f & $<130$ & $<2\times10^{-6}$ &  $<85$ & $1400^{+1600}_{-500}$\\
1115$+$407 &  E      &       0.85       &6.83$\pm0.09$ &130$^{+80}_{-90}$ ($<100^\S$)&1.7$^{+1.0}_{-0.7}$$\times$ 10$^{-6}$&95  &$<$320\\ 
1116$+$215 &  C      &        3.1       &$7.17^{+0.07}_{-0.18}$ & $200^{+110}_{-140}$ ($<80^\S$)   & $7^{+3}_{-5}\times10^{-6}$& 96.5 & $500^{+1100}_{-200}\,^{(\dag\dag)}$\\ 
1202$+$281 &  C      &        2.68      &6.4f & $<$80			& $<2\times10^{-6}$ 		& $<$75 &$<$420\\
1206$+$459 & PL      &        14.9      &6.4f &$<$350&$<6\times10^{-6}$&$<$75&$<$580\\
1211$+$143 &  E      &       0.50       &$6.45^{+0.2}_{-0.07}$   & $40^{+20}_{-30}$& $1.7\pm1.1\times10^{-6}$ 		& 95.1 & $300^{+300}_{-200}$\\
1216$+$069 &  E      &       4.8        &6.4f &$<70$& $<$1.6$\times$ 10$^{-6}$&$<$75&$<$220\\ 
1226$+$023$^\star$ &  E      &       51.05      &6.4f &$<$7&$<$6$\times$10$^{-6}$&$<$75  &$<$17\\ 
1244+026   &  A      &        0.14      &6.65$^{+0.07}_{-0.18}$ & $300\pm200$  ($<150^\S$) & $7^{+5}_{-4}\times10^{-6}$ 	&95.5  &$<$600 \\ 
1307+085   &  A      &       1.19       &6.4f&$<$110      & $<$3 $\times10^{-6}$ 	& $<$75 &$<$920\\
1309$+$355$^\star$ &  A      &       0.67       &6.37$^{+0.05}_{-0.06}$ &180$^{+80}_{-40}$&1.8$^{+0.8}_{-0.4}$$\times$ 10$^{-6}$&99.7  &$<$410\\
1322$+$659 &  C      &       1.08       &6.48$\pm0.09$ &180$\pm110$&2.8$^{+1.6}_{-1.7}$$\times$ 10$^{-6}$&97  &$<$480\\ 
1352+183   &  E      &        1.36      &$6.44^{+0.04}_{-0.07}$ & $150\pm80$            & $4\pm2\times10^{-6}$ 		& 97.7& $470^{+170}_{-300}$\\ 
1402+261   &  C      &        1.4       &$7.47^{+0.07}_{-0.3}$ & $230\pm120$  ($<100^\S$)         & $4^{+2}_{-3}\times10^{-6}$ 		& 96.4& $<$1200\\ 
1407$+$265 &  A      &       41.20      &6.4f &$<$30&$<$1.0$\times$ 10$^{-6}$&$<$75&$<$66\\
1411$+$442 &  H      &       0.25       &$6.43^{+0.05}_{-0.13}$ &250$^{+130}_{-60}$&2.9$^{+1.5}_{-0.7}$$\times$ 10$^{-6}$&99.8  &900$\pm300$\\ 
1415+451   &  E      &        0.4       &$6.35\pm0.07$ & $110\pm80$    &$1.5^{+1.0}_{-1.1}\times10^{-6}$ 		& 97.6 & $700^{+300}_{-600}$\\ 
1427+480   &  C      &        1.6       &$6.40^{+0.09}_{-0.2}$   & $90^{+50}_{-60}$& $1.3^{+0.9}_{-0.8}\times10^{-6}$ 	& 97.1& $<$480\\ 
1440+356   &  D      &        0.58      &6.4f &  $<80$     &$<3\times10^{-6}$ 		& $<$60 & $<$310\\ 
1444+407   &  D      &        1.3       &6.4f & $<$180     &$<2\times10^{-6}$ 	& $<$75 & $1000^{+500}_{-900}$\\
1501$+$106 &  E      &        0.49      &6.4f &$<$60&5$<1$ $\times$ 10$^{-5}$&93  &$<$52\\
1512$+$370$^\star$ &  D      &       8.84       &6.52$^{+0.06}_{-0.08}$ &120$\pm$60&3.9$^{+1.8}_{-2}$ $\times$ 10$^{-5}$&99.5  &$<$240\\
1613$+$658 &  A      &       1.78       &6.4f &$<$60&$<$4$\times$ 10$^{-6}$&$<$75&$<$210\\
1626$+$554 &  D      &       1.46       &6.4f &$<160$&$<7\times$ 10$^{-6}$&$<$75  &$<$190\\
1630$+$377 &  D      &       20.9       &6.48$^{+0.18}_{-0.11}$ &600$\pm$300 &5$^{+3}_{-2}$ $\times$ 10$^{-6}$&99.8  &$<$1400\\ 
1634$+$706 &  C      &        127.9     &6.4f & $<$82      &$<5\times10^{-6}$ 		&$<$75 & $<$680 \\
2214$+$139 &  I      &       0.48       &6.34$\pm0.06$&80$^{+40}_{-40}$&4.0$\pm1.7$ $\times$ 10$^{-6}$&99.9&$<$150\\ 
2303+029   &  C      &        16.2      &6.4f                & $<$ 140      & $<1.6\times10^{-6}$&$<$20 & $<$160\\ 
\hline
\end{tabular}
\end{center}
$^{(\star)}$ Radio loud objects. $^{(\dag)}$ In this source only an iron  line with a broad profile was significantly detected. $\,^{(\dag\dag)}$  The detection  of   the  {\tt LAOR} line at  6.4~keV   is
significant, i.e. F-test $>$~95~\%. $^{(a)}$ Best fit model: A: blackbody; B: multicolor blackbody; C:
bremsstralung emission; D: power law; E: double
blackbody; F: {\tt absori} model ; G: double {\tt absori} model $+$
blackbody; H: partial--covering
(cold) $+$ Raymond-Smith; I: partial--covering(warm)$+$emission line; J:
absorption $+$ blackbody; PL: simple power law.  Models with $^{(\ddag)}$ include an
additional cold absorption component.  All but one type of models
(i.e.  model PL) also include a power law component accounting for the
hard X--ray band (see \citet{pgpaper1}). The EW labeled with $^{(\S)}$
correspond to the neutral iron line added in addition to the ionized one.
\end{footnotesize}
\end{table*}

\begin{table*}
\caption{Broad iron line properties. The table includes the name of
the QSO for which a broad iron emission line  is significant. The line
was modeled by a {\tt LAOR} line  assuming a fixed rest frame
energy of the  line to 6.4~keV, an  inner (outer) radius  of 1.26(400)
$R_g$, an inclination of 30$^\circ$, and $\beta$ equal  to -2 (see text for
further details). The intensity and  rest--frame equivalent width  of
the broad line, as well as the F--test probability, are also given.}
\label{tab:broad}
\begin{footnotesize}\begin{center}
\begin{tabular}{c c c c }
\hline \hline \\ 
PG Name   &Intensity &EW$_{broad}$ &  P(F--test)\\ 
& ph cm$^{-2}$s$^{-1}$ & eV &  \%\\  \hline\hline\\
0007$+$106 &$(1.5\pm0.7)\times10^{5}$& 220$^{+110}_{-100}$  &99.2 \\
0050$+$124 & $(6\pm3)\times10^{-6}$& 100$^{+40}_{-50}$  & 99.6\\ 
1116$+$215 & $1.7^{+1.7}_{-1.2}\times10^{-5}$& 500$^{+1100}_{-200}$ & 97.6\\
\hline
\end{tabular}
\end{center}
\end{footnotesize}
\end{table*}

\begin{figure*}
   \centering
   \includegraphics[angle=0,width=6cm,height=5.5cm]{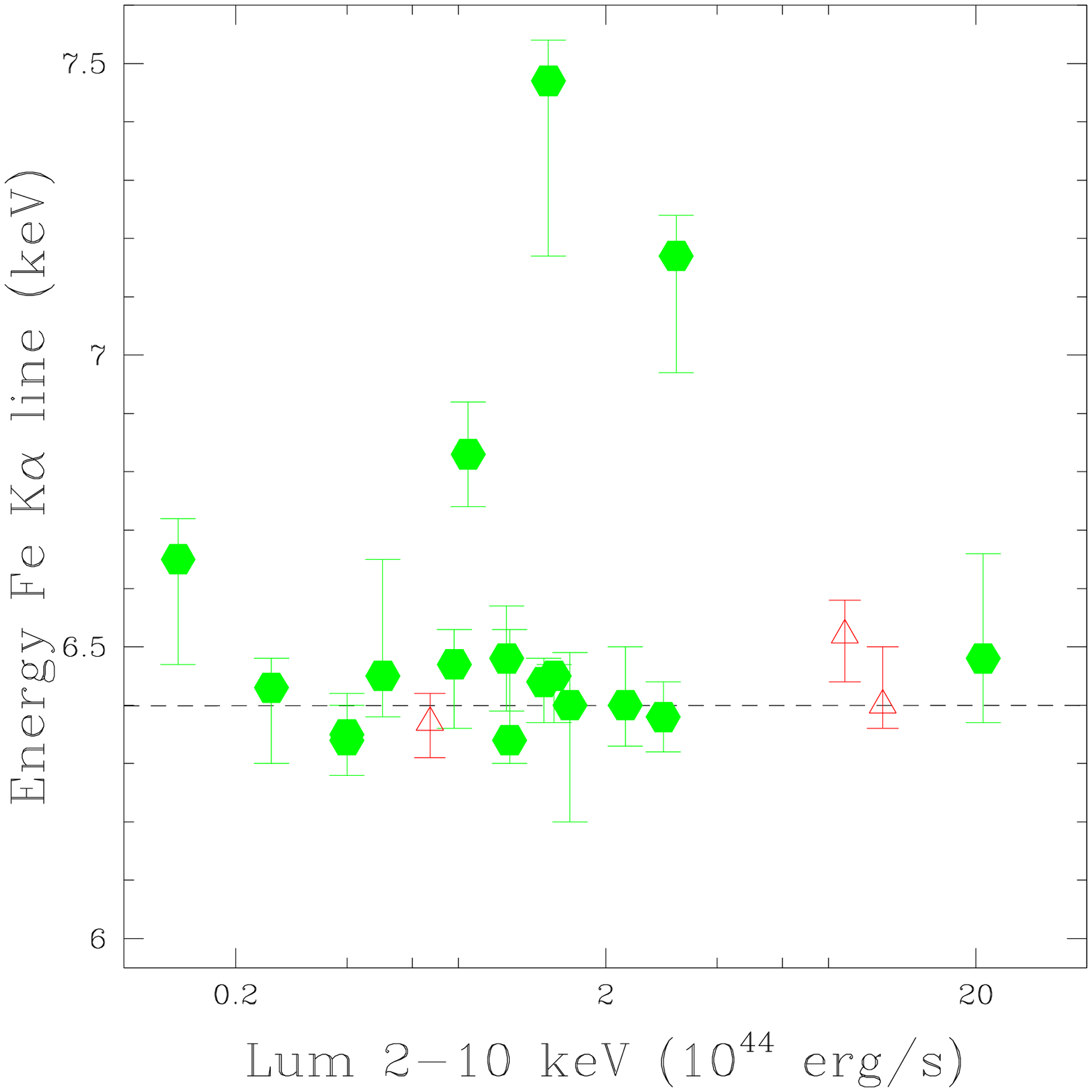}\hspace{1cm}
   \includegraphics[angle=0,width=6cm,height=5.5cm]{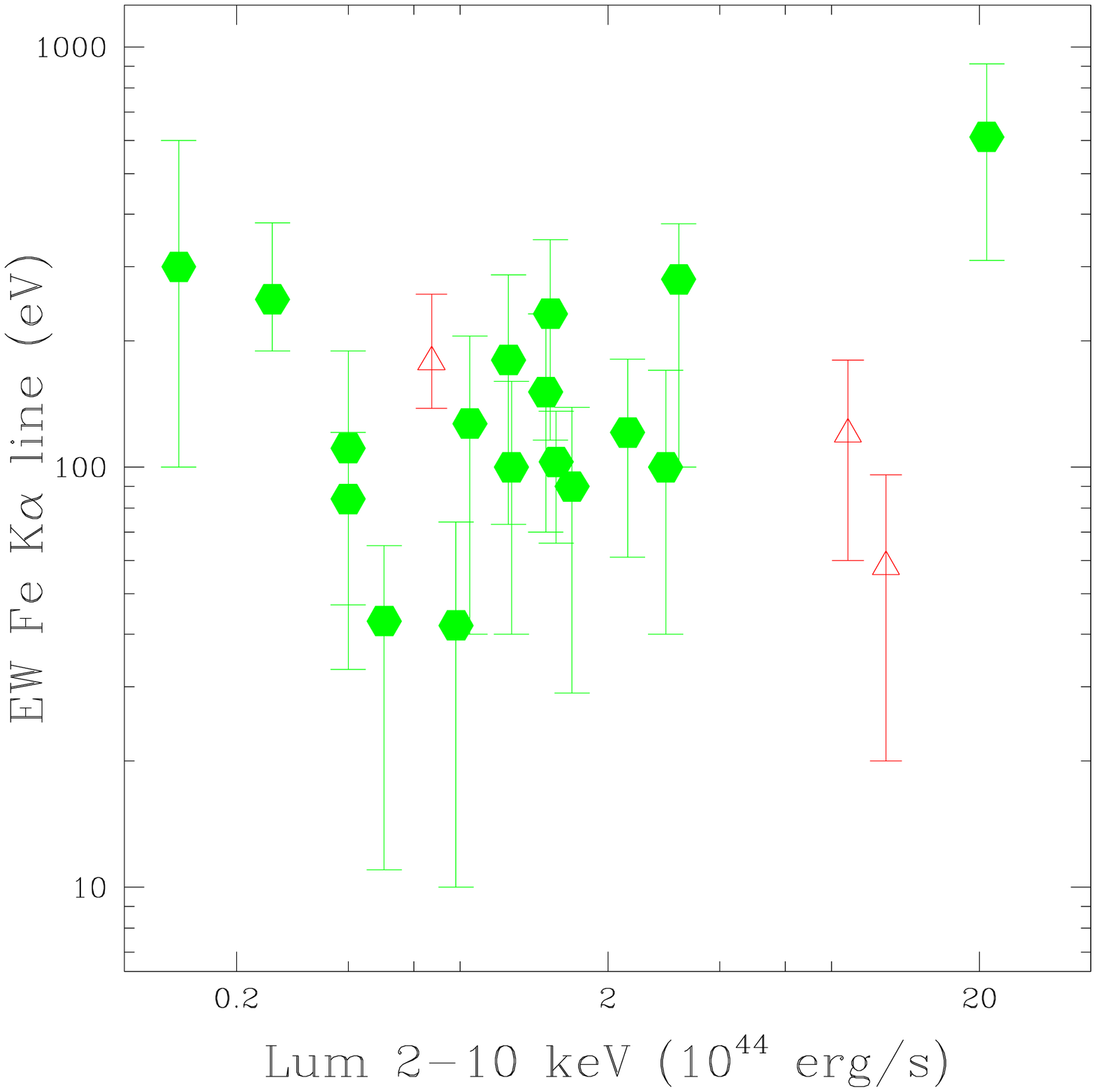}\hspace{1cm}
   \caption{(a) Centroid Fe K$_\alpha$ line as a function of the
\lum. The radio--loud QSOs are plotted as empty triangles.
(b) Equivalent width of  the narrow line significantly
detected, i.e. F--test $>$95\%,
as a function of  the \lum. The  radio--loud  QSOs are also  plotted as
empty triangles.  }
\label{fig:fe_lum} 
\end{figure*}

Fig.~\ref{fig:fe_lum}a shows  the  centroid  of the iron  fluorescence
narrow emission line plotted as a function of 2--10 keV luminosity
for the   20 QSOs with  significant  detection of such an emission
feature (see Table~\ref{tab:iron}). There  is  no apparent trend  with
luminosity or radio--loudness.      The  centroid energy  of  the   Fe
K$_\alpha$ line in  all but four quasars (1115$+$407 1116$+$215
1244$+$026 1402$+$261) corresponds to low  ionization states, i.e.  Fe
I--XVII
\citep{makishima}.   Fig.~\ref{fig:fe_lum}b shows  the EW  of Fe
K--shell narrow emission lines as a function  of luminosity in the
hard  band; however; however  no obvious   correlation   emerges from this   plot (see
Sect.~\ref{sub:baldwin}).\\

Besides the ubiquitous presence of a  narrow Fe emission line centered
at $\sim$ 6.4 keV, recent  works  based on Seyfert~1 samples  observed
with  both the   \chandra~\citep{yaqoob}   and     \xmm~\citep{pounds}    high
signal--to--noise  spectroscopic  data have  found   in  some
objects evidence of additional underlying  broad and skewed emission supporting
the original {\it    ASCA}   claims for a    relativistically  broadened
component \citep{nandra}.  Consequently, we added a disk line to the fitting model
 to account for   any such ``broad'' component.  For  this we used the {\tt LAOR} model following recent
\xmm~results that  advocate a spinning  black hole in nearby Seyfert
galaxies \citep{wilms,fabian02}.  In our  fit we assumed a  fixed rest
frame  energy of  the line to  6.4~keV, a  power  index of  the radial
emissivity law  $\beta$   equal to  -2, an  inner  (outer) radius   of
1.26(400) $R_g$, and  an inclination of 30$^\circ$~\footnote{These are
the   same  parameters used  in   \citet{porquet}  for  the   broad Fe
K$_\alpha$ line  observed in the  \xmm~spectrum of the luminous quasar
Q0056$-$363}.  The resulting  EW of the  broad component  is listed in
last  column  of  Table~\ref{tab:iron}.   Introducing  this  spectral
feature   significantly improved  the  fit    statistic in  3  sources
(0007$+$106,  0050$+$124, and 1116$+$215).  As  a further test we used
the {\tt GAUSS} model with $\sigma$ free to vary,  instead of the {\tt
LAOR}  model,   and iron  lines with  a  broad  profile resulted being
detected in  the same objects.   The intensity, equivalent width,
and the F--test for each line are  shown in Table~\ref{tab:broad}.  In
0007$+$106,  an iron fluorescence emission  line  with a broad profile
was  detected without the additional  presence of a narrow core, while
in  the last two,  0050$+$124 and 1116$+$215, both  the narrow and the
broad emission feature are present.   This limited statistic, however,
does not allow more detailed investigation of the line parameters.

For the QSOs with a significant Fe K$_\alpha$ emission line  we also
tested a model with a neutral reflection component  \citep[{\tt
PEXRAV} in XSPEC, see][]{magdziarz}.  
This parametrization led to no significant improvement in  goodness
of  fit in the  vast majority of  those sources with resulting values of
$R=\Omega/$2$\pi$  (i.e. the solid  angle  subtended  by  the cold
reflecting  material to the X--ray source  located above it) basically
unconstrained  but  consistent  with zero. As   expected, the bandpass
(0.3--12 keV)  and the limited statistics  usually affecting  the high
energy ($E$\simgt~8~keV) region  of the \epic~spectra are not suitable
to constrain the strength of this bump--like emission component that peaked
at $\sim$ 40 keV \citep{lightman}.  Nonetheless, in two cases,
0007$+$106 and 1309$+$355,   introduction  of a  reflection
component significantly improved the fit  at $>$ 98\% confidence level
with  a resulting       $R$    = 2.6$^{+1.4}_{-1.2}$  and    $R$     =
2.9$^{+1.5}_{-1.1}$,  respectively.  Such   large  values  of $R$  
indicates that simple slab geometry for  the reprocessing medium is
inadequate, compared to a  partial--covering scenario  with an additional spectral
component emerging  at high energies  or an anisotropic geometry,
which offer more likely explanations. The iron line parameters are
not affected when the reflection model parametrization is applied.



\section{Discussion}\label{sect:discussion}

Our study reveals that  about 50\% of the  QSOs in the sample
show   a    significant iron  fluorescent    emission  line   in their
spectra. Iron K--shell emission therefore is a common feature in
the X--ray spectrum of   optically--selected QSOs. In  particular,  we
detected 20  Fe K$_\alpha$  lines with a  narrow   profile and 3  broad
lines.  In  the following section we   focus on  properties of the
former feature and in Sect.~\ref{sec:broad} discuss characteristics
of the broad emission lines.

\subsection{The narrow iron line}\label{sec:narrow}

The  energy centroid   of     all   but   four  narrow     lines
(Fig.~\ref{fig:fe_lum}a) is consistent with matter in low ionization
stages  (FeI--XVII).  The    mean value   of  the energy  for  the 20
significant lines detected  is  $6.53\pm0.09$~keV  with  a  dispersion
lower than 0.3~keV. When only the 16 detected low ionized Fe lines are
considered, the  value  found  is  $6.41^{+0.03}_{-0.04}$~keV   with a
dispersion lower than  0.05~keV. This best  simultaneous estimate
of the energy of the line and its intrinsic spread was calculated
using the maximum likelihood technique of
\citet{maccacaro}. The errors were obtained through the 68\%
confidence contour level. This finding turns out  to  contradict
the {\it ASCA} results of
\citet{reeves},   who  
found \simgt~50\% of the detected Fe lines at energies $>$6.4 keV in a
sample  of QSO with  similar luminosities  \citep[see also][]{george}.
Nevertheless, considering only    the common objects  in  both samples
(i.e.   0050$+$124,   1116$+$215,   and 1226$+$023)   the  only  minor
discrepancy is found in  the case of  0050$+$124 where an ionized line
was only marginally  detected in the ASCA  data. The  line detected in
1226$+$023 is not   significant in none of  the  spectra, and the  two
centroid   energies of 1116$+$215 obtained   with  {\it ASCA} and {\it
XMM--Newton} are compatible within the errors.  On the other hand, our
results are  compatible with other recent  \xmm~studies of  QSOs; in a
sample of five  high luminosity RQQs,  \citet{page_rqq} find that  the
detected iron emission  lines are mainly neutral.  In a larger  sample
analysis of low--z QSOs performed with \xmm,
\citet{porquet_21} detect an iron line in twelve out of 21 QSOs, and the
values of the centroid of the Fe lines agree in  all but two sources.
Our finding  also matches well  the  properties of iron   K$_\alpha$
emission in Seyfert 1 galaxies  observed with \xmm~and \chandra,
where  the     mean  peak   of   the narrow    line    is  observed at
$\langle$E$_{K_\alpha}\rangle \approx$ 6.4 keV
\citep{nandra,yaqoob}.

The    emission  lines with    energy E$_{K_\alpha}>$   6.4  keV in
1115$+$407 and   1116$+$215 are both consistent   with emission due to
iron  at  very high ionization   stages   (i.e.   He--and   H--like,
respectively); while  in the case  of the line observed in 1402$+$261,
the  observed energy exceeds any ionization  states for the K$_\alpha$
emission.  Reeves et al. (2004) have explained the origin of this
feature as a blue--shifted line originating in an accretion disk with 
high inclination.

The intensity   of  the significantly   detected  narrow lines appears to  be
independent of  the X--ray luminosity for L$<10^{45}$\ergs, (Fig.~\ref{fig:fe_lum}b;
see the next Section for a more detailed  discussion), with an average
value  $\langle$EW$\rangle$ =  80$\pm20$~eV  and an intrinsic
dispersion $\sigma_{EW}<$~40~eV. Such a  value is lower than
163$\pm$17, i.e. the  $\langle$EW$\rangle$  inferred by  \citet{reeves}
for the radio--quiet QSOs in their sample.  Nevertheless, the values
for the EW of the detected iron lines are consistent with the
measurements of \citet{page} who using \xmm~data of five QSOs
 find EW values within the range  30--180~eV.  The result is  also  compatible with the  typical
values found for Seyfert galaxies; the mean EW obtained by \asca~for a
sample  of Seyferts was  114$\pm$10~eV \citep{nandra}, and  more recently,
high resolution   \chandra~observations measured  a  weighted mean  of
65~eV for the narrow cores \citep{yaqoob}.\\

The average properties  of Fe K--shell  narrow emission lines observed
for our objects are therefore  very similar to  those inferred for the
low--luminosity AGNs. Consequently, the origin of this feature in QSOs
and Seyferts is  expected to be the  same.  It is widely accepted that
the fluorescence  emission line emerges  in `cold' matter distant from
the central X--ray source consistent to its energy and narrow profile.
Even  if the  emitting   medium  is located   in  the optically--thick
accretion disk,  regions near the  radius of marginal stability  for a
not--rotating or  spinning black hole  (i.e.  6 and 1.24 gravitational
radii, respectively) are ruled out.    Furthermore, zones of the  disk
very near to the X--ray source are likely to be ionized given the high
irradiation.  A substantial fraction of the line  flux could be due to
reflection of the primary  continuum off the  inner walls of the torus
into our   line  of  sight  \citep{ghisellini94}.   This parsec--scale
structure is  a key component of  AGN Unification models, and  is
indeed thought to  subtend a large solid angle  to the  X--ray source
\citep{matt_2000}.  Assuming standard values for Type 1  AGNs (i.e.  a
torus column density
\nh~$\sim$10$^{24}$ \cm2,  a viewing angle\simlt~20$^\circ$, and a half
opening angle of the torus $\theta \sim$ 30$^\circ$), the predicted EW
of   the     line   produced  by  a    torus  with solar abundances   is  $\sim$    100 eV
\citep{ghisellini94}, in agreement with our observations.

Finally, we should consider    the possibility of a  contribution,  at
least partially,  from the optical--to--ultraviolet Broad  Line Region
(BLR) to the   line emission.  On the  basis  of  variability  and
emission properties (energy, width, EW),
\citet{yaqoob01} inferred that the  bulk of the  iron line in NGC~5548
is  indeed  consistent  with   emission  in  the  outer   BLR.   Such
interpretation has been  also  proposed by  \citet{kaspi} for  the  Fe
K$_\alpha$ line found in   a very long \chandra~HETGS   observation of
NGC~3783. The  equivalent hydrogen  column  of the BLR, N$_H$,  can be
estimated using the EW of the iron line, the  index of the power law,
and  the  fraction  of sky  covered  by the  clouds  \citep{yaqoob01}.
Considering the mean  index of the  power law obtained in analysis
of  the  QSOs in  the  sample, $\langle\Gamma_{2-12\,keV}\rangle=1.87$
\citep{pgpaper1}, and the $\langle$EW$\rangle$=80~eV, we get

\begin{center}
\begin{equation}
N_H\simeq 2.5\left(\frac{0.35}{f_c}\right)\times 10^{23}\,cm^{-2}
\end{equation}
\end{center}

\hspace{-0.7cm} where $f_c$ is the covering factor. Assuming spherical
covering, i.e.  $f_c=1$, the   equivalent hydrogen column  density is
$N_H\simeq9\times10^{22}$~cm$^{-2}$. This value can be considered as a
lower limit to the absorption column and is, therefore, in agreement
with typical values of BLR clouds, i.e. $N_H\simeq10^{23}$~cm$^{-2}$.

Emission lines from ionized iron (i.e. with an energy $>$ 6.4 keV) can
be   produced      by    reflection in   an ionized       accretion  disk
(e.g. \citet{reeves01}) or, alternatively, by warm scattering material
along the line of sight \citep{bianchi}.  The latter mechanism also appears
likely   in the case of  1116$+$215, for which  the  presence of a
warm medium is suggested by the  detection of warm absorber signatures
in    the  soft X--ray    portion  of  the  spectrum \citep{pgpaper1}.
Similarly,  the strength of  the Fe  line  in 2214$+$139 is consistent
with an  origin  in the  low--ionization component of  the two--zone
warm absorber observed in this source \citep{piconcelli}.

\subsection{The broad iron line}\label{sec:broad}

In contrast to the  narrow Fe K$_\alpha$  emission line detected in 
20 of
the sources, the  broad component seems to  be less common in QSOs with
only 3 out of 38 significant detections.

Emission from  an accretion disk reaching down  to a few gravitational
radii  of  the  black  hole is the  most  likely  explanation for iron
K$_\alpha$ lines with a broad profile.   Such a skewed velocity profile
indeed suggests that fluorescence takes place  in a strong gravity
regime \citep[see][for a review]{fabian00}.  Except for  a very  few remarkable
cases  \citep{porquet,comastri},  most of the
detected   relativistic iron   lines  have  been    detected  in nearby
Seyfert--like AGNs  \citep{nandra}. Their   presence  in QSOs
appears to be rare, which  has been  interpreted as evidence of
different physical properties of the  accretion disk in these objects,
such as high ionization or  small transition radius to radiatively
inefficient accretion flows. Nevertheless    a selection  effect  due   to  the poor
statistical quality  of most X--ray quasar  spectra  must be  borne in
mind, and detection  of broad Fe  lines in three  quasars of our sample
therefore appears noteworthy.  

The broad features observed in 0007$+$106 and 0050$+$124 could also be
explained as contributions from  different iron lines
at different ionization  levels as suggested by  the presence of  some
positive data--to--ratio  residuals  in the  range 6--7 keV  
\citep[see also][]{gallo}.  However,  current \xmm~data does  not allow us to
 investigate    this   hypothesis further.   Similarly,   the  limited
signal--to--noise of the observation of  1116$+$215 hampers a detailed
analysis of the  broad iron lines and affects  the significance of the
inferred spectral parameters, i.e.   leaving almost  unconstrained  the
value of the line intensity ( EW=$500^{+1100}_{-200}$~eV).

\subsubsection{Behavior of the EW with L$_{2-10\,{\rm keV}}$}
\label{sub:baldwin}

\begin{figure*}

   \centering
\includegraphics[width=6cm,angle=90]{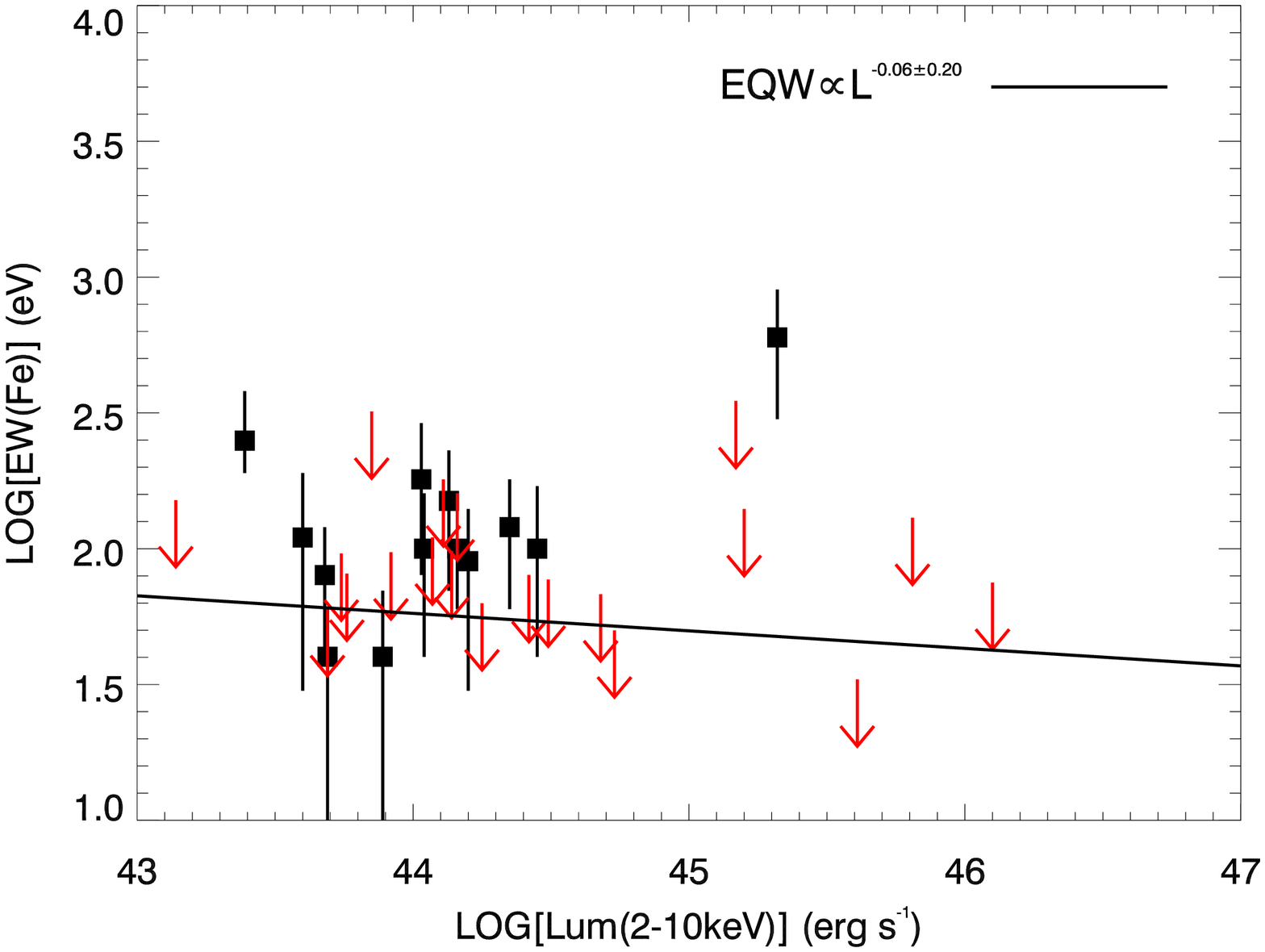}
\includegraphics[width=6cm,angle=90]{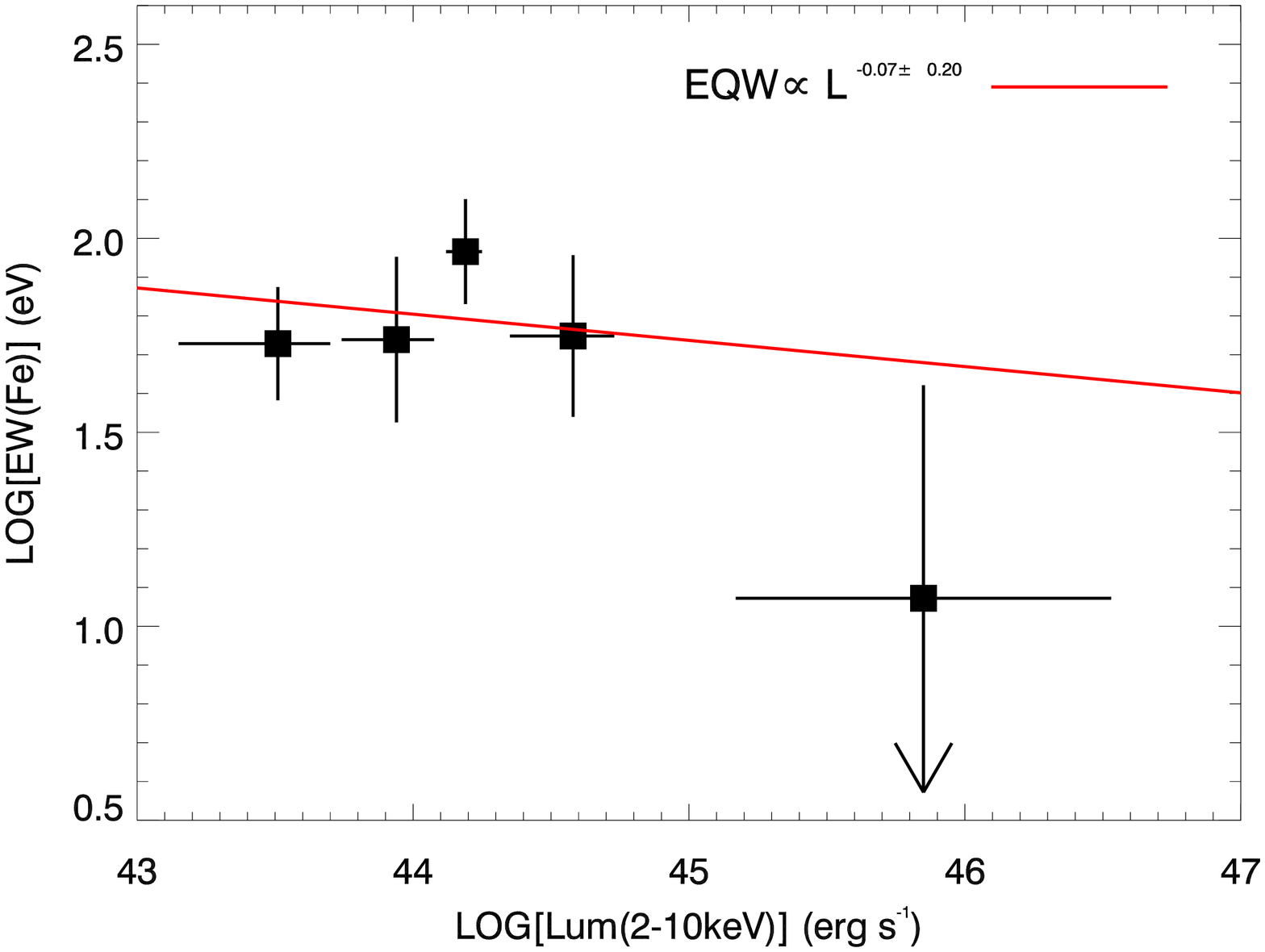}

    \caption{Relation between the  equivalent width of the Fe K$_\alpha$
   line and the \lum~for the radio quiet  quasars. (Upper panel) The plot
includes  measurements for each set of data. The solid line shows the fit
to all the detections while for the dashed line only  the significant
detections are considered.(Lower panel) The plot represents all measurements
binned in five groups. The solid lines show the
   fit to   the  five points.  } 
   \label{fig:bw} 
\end{figure*}

The Baldwin effect was first reported in the optical and ultraviolet emission lines
of QSOs. \citet{baldwin} observed a decrease in the equivalent width of
the  C$_{IV}$ emission line   with increasing  ultraviolet luminosity.
This anticorrelation between  the luminosity and the equivalent  width
was also  found in the X-ray  band for the Fe K$_\alpha$  emission
line. \citet{iwasawa} studied 37 AGNs observed with
\ginga~and found the relation EW(Fe K$_\alpha)\propto L_{2-10\,keV}^{-0.20\pm0.03}$. A similar study of 39
AGNs performed with \asca~also confirmed this effect
\citep{nandra}. More recently,  \citet{page_baldwin} obtained  a relation  of EW(Fe
K$_\alpha)\propto   L_{2-10\,keV}^{-0.17\pm0.08}$  for a  sample of  53
Type~I  AGNs observed with \xmm.

In Fig.~\ref{fig:bw}, we show     the relation between    the 2-10~keV
luminosity and  the equivalent width of the  narrow Fe K$_\alpha$ line
for the objects in our  sample.  In order to  avoid mixture with RLQs,
where   hard  X--ray  emission could   be  {\it contaminated}   by the
relativistic jet components,  we have considered  here only RQQs.

Several different techniques have been compared  to study the behavior of
the EW with luminosity.

First,  linear regressions  of $\log{{\rm  EW}}$  versus  $\log{{\rm
L_{2-10\,keV}}}$ were   performed  using  ASURV ({\it   Astronomy
Survival Analysis}, Fiegelson  \& Nelson, 1985) package.  ASURV facilitates
a correct  statistical   analysis when censored data   (upper
limits in this case) are present, although errors are not considered in
the calculations   (see   below).  This   allows   us  to  study   the
relationship between EW  and $L_{2-10\,keV}$, including all significant
and  non--significant  (the upper   limits) detection  of  the neutral
narrow lines.   For the four ionized  lines found in the  analysis, an
extra   neutral component was added    to  the fit  model.  This
component, modeled by a Gaussian  line with $\sigma$=0, is not
significantly required  in any   of the four sources.   The  upper limits of  the
corresponding EWs can also be found in Table~\ref{tab:iron}.

Within the ASURV  package, we   performed   a Spearman--Rank  test
considering the detections and    upper limits, and found only a 30\%~probability that  $\log{{\rm EW}}$ and $\log{{\rm L_{2-10\,keV}}}$ are
correlated.  The fit calculated with the Buckley--James
method gives  a    power law index  of   $\alpha=-0.06\pm0.20$,  where
EW(FeK$_\alpha)\propto  L_{2-10\,keV}^{\alpha}$.  The result  obtained
by~\citet{page_baldwin}  with    the     same           method,
$\alpha=-0.17\pm0.08$, is compatible with our value within the errors.

The disadvantage  of using the  ASURV package in  the data analysis is
that  actual   detections are  considered  without taking  errors into
account,  thereby underestimating the  uncertainties in determining the
slope in certain cases.  In order to avoid this problem, we calculated
the mean EW in five different luminosity bins with a similar number of
objects in each bin, therefore we  studied the relationship between EW
and $L_{2-10\,keV}$ using these five bins.  The weighted mean EWs were
calculated considering asymmetric errors statistics (Barlow 2004). The
index of the power law obtained is $\alpha=-0.07\pm0.20$, very similar
to the results given by ASURV when  upper limits are considered in the
fit. Fig.~\ref{fig:bw}(b) shows the data together with the fit.\\

\begin{figure*}

   \centering
\includegraphics[width=6cm,angle=90]{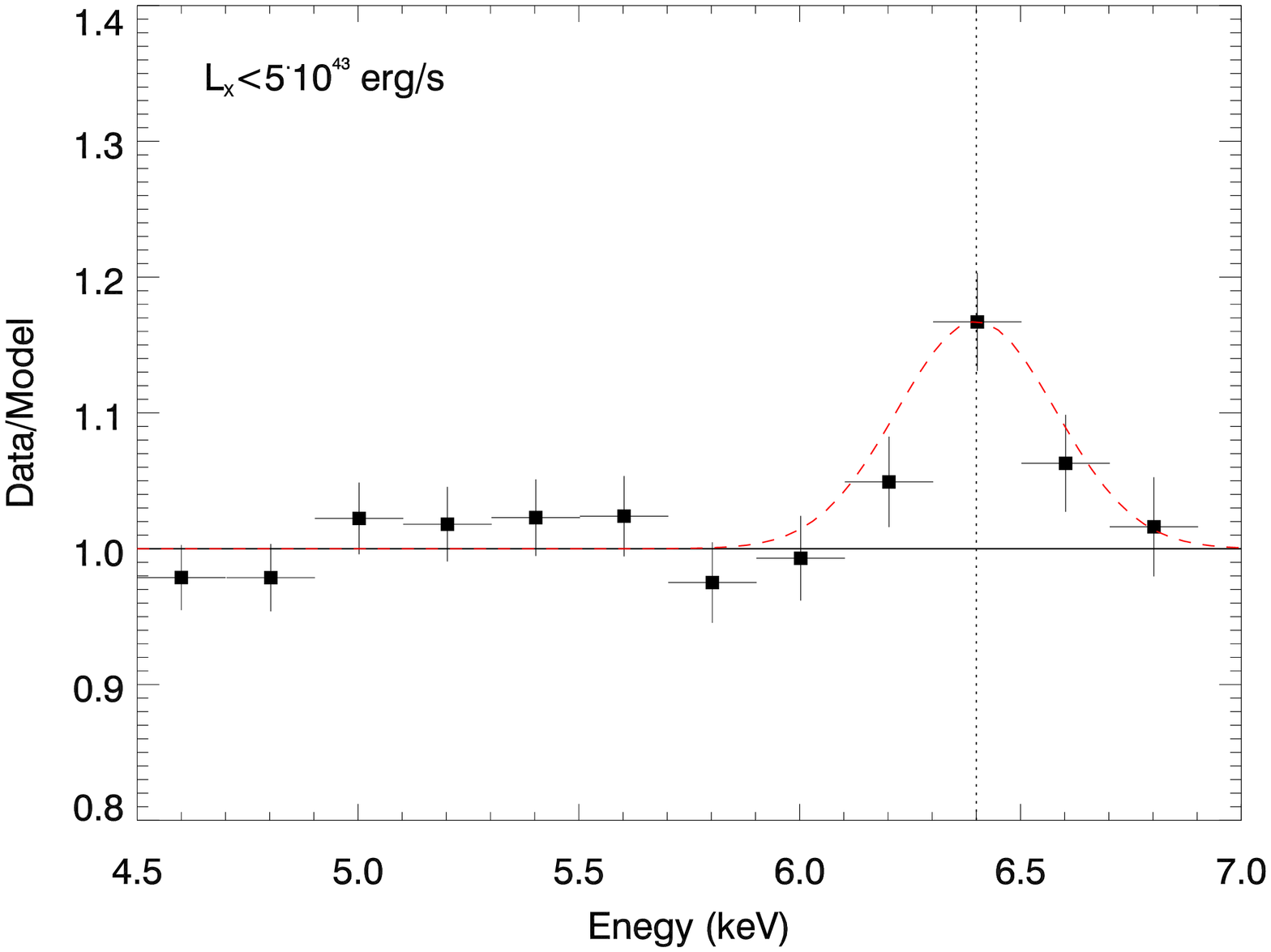}
\includegraphics[width=6cm,angle=90]{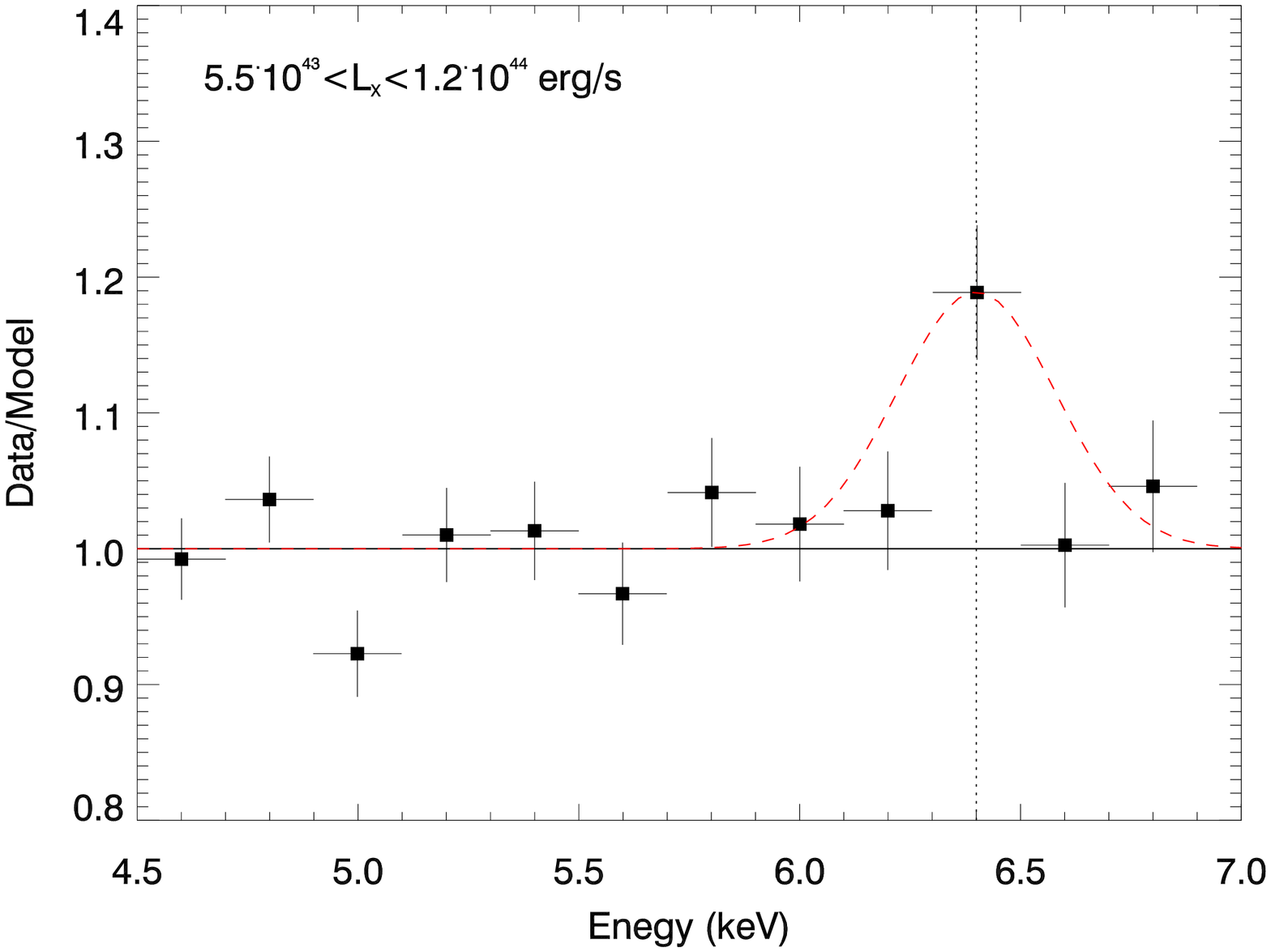}
\includegraphics[width=6cm,angle=90]{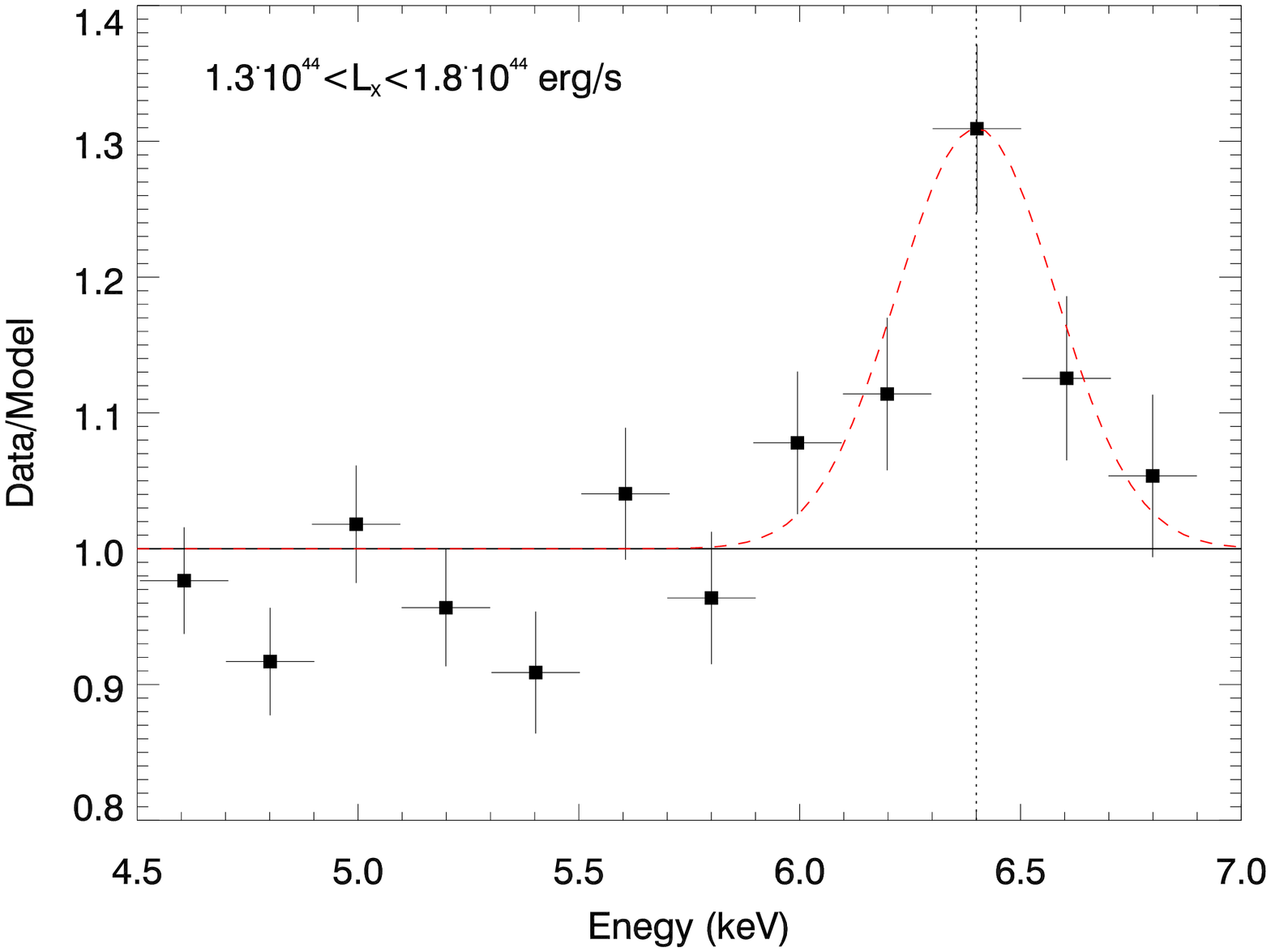}
\includegraphics[width=6cm,angle=90]{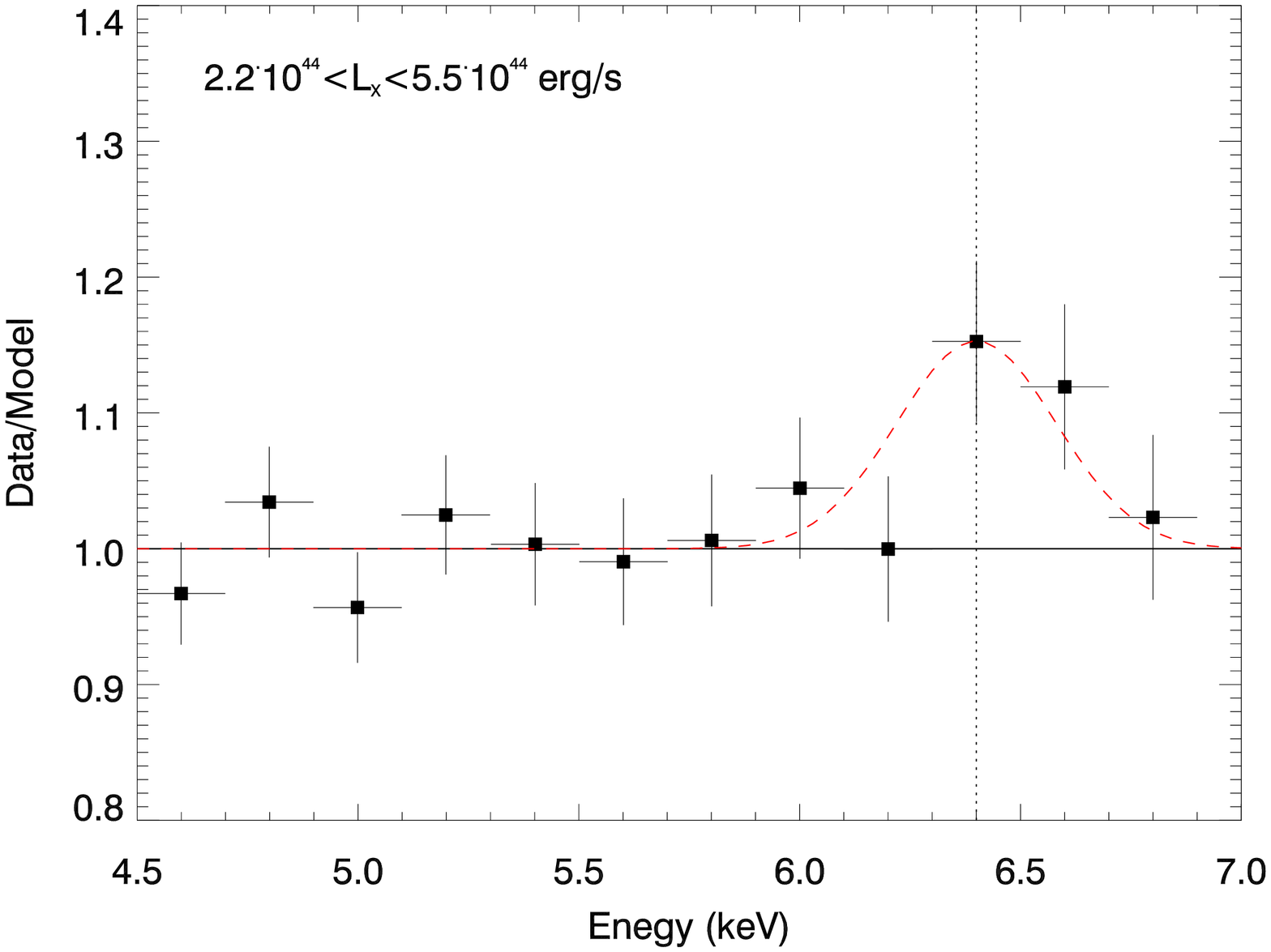}
\includegraphics[width=6cm,angle=90]{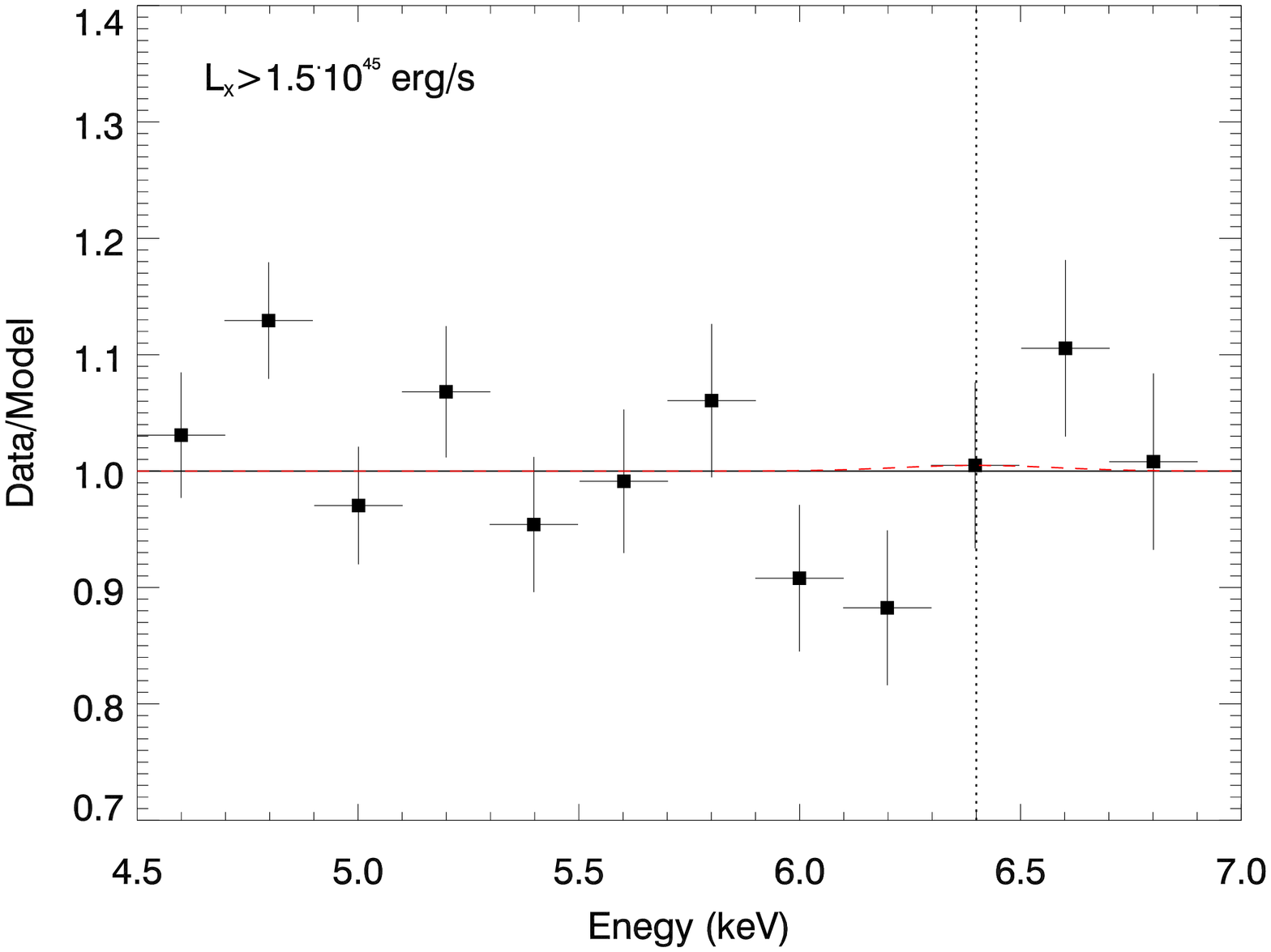}
\caption{Mean data--to--model ratios in five luminosity bins,
assuming for each QSO  the best fit model and  excluding the iron line
and correcting by redshift. The observed lines represent a Gaussian
profiles with centroid energy 6.4~keV, and width equal to the
intrinsic \pn~resolution, normalized to the 6.4~keV data point.}
\label{fig:bins}
\end{figure*}

Fig.~\ref{fig:bins}  shows  a   comparison of  the data--to--continuum
residuals  in the  4.5-7~KeV  energy band,  obtained by  combining the
ratios  corresponding to   objects  selected  in five   luminosity
intervals.   For each interval, we  considered the best fit model
of  each  QSO  but  excluded  the  iron   line in  cases where  it is
significantly detected.  The   unbinned data--to--model ratio  of each
QSO is translated into the rest   frame and grouped in bins  of
0.2~keV  size.  Each plot   also  includes a  Gaussian emission   line
broadened to the intrinsic resolution of the  {\it pn} at  the energy
of  the   iron line,  i.e.   FWHM=150~eV.  The   plots show residuals
compatible with the neutral iron line  being narrow.  The residuals in
the last bin are larger and  randomly distributed.

An anticorrelation between  EW and  hard X--ray luminosity is   also
compatible with our  data. However, other behaviors  different from a power law
cannot be   ruled out. The data actually suggest a constant EW for
$L_{2-10\,{\rm  keV}} \simlt 10^{45}$~\ergs and a stepwise
decrease above this threshold. 
As    can  be   seen   in Fig.~\ref{fig:bw}(b) the anticorrelation
between EW and $L_{2-10\,{\rm  keV}}$ is actually determined mainly by the
highest luminosity bin.\\

The physical origin of variation in the iron line equivalent width
with luminosity is related to where the line is originated.  Assuming
that the Fe K$_\alpha$ line is produced by fluorescence, various
scenarios have been proposed.  \citet{iwasawa} suggested that the
emission line is produced in the clouds of the Broad Line Region.  The
X-ray primary emission ionizes these clouds explaining the observed
decrease in the equivalent width with luminosity.  \citet{nandra}
proposed that for iron lines, which are most likely produced in the
accretion disk, an increasing degree of ionization with luminosity
could also account for the Baldwin effect.  Assuming the emission line
is originated in the torus, \citet{page_baldwin} suggested that
increasing X-ray emission could cause a decrease in the covering
factor of the torus and hence a decrease in the equivalent width of
the iron line. A step function behavior of the value of EW with
luminosity favors our explanation in terms of the increase in the
degree of ionization suggested by \citet{iwasawa}.  When the material
reaches a particular ionization state, the emission line could be
suppressed by resonant trapping and/or full electron stripping and the
EW would then abruptly decrease. In such a case, the material where
the line is emitted should be most likely identify with the BLR or accretion
disk. In the last case, the average unresolved profile
indicates an origin in the outermost regions of the accretion disk.\\

Nevertheless, our  analysis    cannot rule out  a   smooth  power  law
dependence, EW(FeK$_\alpha)\propto     L_{2-10\,keV}^{\alpha}$,     as
suggested by \citet{page_baldwin}.  The slope  of the power law ranges
between -0.27 and 0.14, taking into account the different analysis and
their corresponding uncertainties.  More highly luminous QSO with high
signal--to--noise need to  be added to  this analysis to reduce  large
uncertainties in determining  the slope.   A possible divergence  from
previous works could be due  to a selection  effect, as they claim the
presence of  X-ray Baldwin effect and   include both RQQs  and RLQs in
their  sample.   In  particular,   most  of  the  objects  with   high
luminosity, $L_{2-10\,keV}>10^{45}$~\ergs, were  radio loud.  However,
as shown by
\citet{reeves},  there   is  a  strong  anticorrelation   between iron
K$_\alpha$  EW and radio loudness.   This  hypothesis is also strongly
supported  by anticorrelation between  radio loudness and the strength
of the  soft excess component  studied in \citet{pgpaper1}. This trend
could be  due to  the likely presence   of an additional  jet emission
component that  weakens the strength  of the  iron line.  Inclusion of
RLQs in the highest 2--10 keV luminosity bins  in previous works could
therefore have affected the results on the  X--ray Baldwin effect. The
choice of our sample minimizes this bias.

A similar analysis of the radio  loud objects is  hampered due to the
low  number of  RLQs    in the  sample.  Furthermore,   a  significant
Fe~K$_\alpha$ line was detected in only one out of five RLQs.



\section{Summary and Conclusions}

We present here an  analysis  of the   K$_\alpha$
fluorescent iron emission line in the \xmm~spectra of  38 PG~QSOs, including  the
presence  and properties of emission lines with both
narrow and broad profiles.  \\

A  narrow unresolved   (i.e.  FWHM$\leq7000$~km~s$^{-1}$) emission  Fe
K$_\alpha$ line  was detected  in $\sim$  50\% of the  objects studied.
The energy  centroids indicate that   the majority  of the  lines 
originated  in cold material (Fe  I--XVII). The  mean equivalent width
for the  narrow line is  $\langle$EW$\rangle=80^{+30}_{-20}$~eV with a
dispersion  $\sigma_{EW}<$~40~eV.  Such  properties  are  consistent with
several scenarios for the origin of  the line. Fluorescence likely
occurs  in the outermost parts of   the accretion disk  and/or in even
more external  regions, such as the  molecular torus or the  clouds of
the Broad  Line  Region.   For two   objects,  i.   e.   1115+407  and
1116+215, the line  energies are associated  with the highest ionization
stages of iron, suggesting emission from an  ionized accretion disk,
due to scattering by warm material located in the line of sight or due
to the presence of an outflow.\\

We also investigated the presence of an X--ray  Baldwin effect for the
Fe K$_\alpha$ line of the RQQs. The RLQs were
excluded  in  order  to   avoid   possible  contamination    from  a
relativistic  jet component in   the X--ray emission, which  could have
biased the anticorrelations published  in the literature prior to  our
study. However, no clear correlation was found between  strength of the
iron line and  2--10~keV luminosity.  Assuming a power
law   behavior   EW(FeK$_\alpha$)$\propto    L^{\alpha}_{2-10\,keV}$,
$\alpha$ turned out to  in the range between -0.3 and 0.14.   \\

Broad   relativistic lines were detected   in three objects, less
than  10\%   of the QSOs  in  the sample.   These objects have  low
luminosities    (L$_{2-10\,keV}\simlt  10^{44}$~\ergs) suggesting    a
possible anticorrelation between  the  X--ray  luminosity (hence the
ionization   state of the uppermost  disk layers) and the  strength of 
lines      originating   in     the   disk,      as has been   proposed    by
\citet{nayakshin}.  However, a selection   effect due  to  limited
statistics affecting the    most distant (very   luminous)  QSOs
prevents us from drawing firm conclusions on this point.\\

After comparing   our  results  with those   of recent  \xmm~and
~\chandra~observations of    Seyfert galaxies,    it seems   that 
characteristics of  the Fe  K--shell  emission lines detected  in  the
spectra do not  significantly change  as a function of the X--ray
luminosity.  We report average values of  the EW and the energy
centroid  of the line which are both  consistent with those in Seyfert galaxies.
Moreover,  \citet{pgpaper1} also found a similarity between the
properties  of  the cold  and warm   absorption  in QSOs   and Seyfert
galaxies.   Our analysis therefore allows an extension of the well studied
observational  properties of the  nearby  X--ray  bright
Seyferts to the QSO realm, at least for those optically selected.


\begin{acknowledgements}
The authors would like to thank the referee, Dr. James Reeves, for this
encouraging report and useful  comments, which  
improved the paper significantly.  We also thank  the \xmm~SOC  science support team
members.  This paper  is based on observations  obtained with \xmm, an
ESA science mission with instruments and contributions directly funded
by ESA Member States and the  USA (NASA).  This  research has made use
of the NASA$/$IPAC Extragalactic  Database (NED) operated by
the Jet Propulsion   Laboratory, California Institute   of Technology,
under contract with the National Aeronautics and Space Administration.

\end{acknowledgements}

\end{document}